**Terminological and methodological discrepancies concerning the radionuclides' effective, environmental and biological half-lives**

Grzegorz Oloś[1], Agnieszka Dołhańczuk-Śródka[1]

[1]Institute of Environmental Engineering and Biotechnology, Faculty of Natural Sciences and Technology, Opole University, Kominka Street 6, 45-032, Opole, Poland

tel. +48 77 401 60 50, +48 77 401 60 51

Corresponding author: Grzegorz Oloś, golos@uni.opole.pl

ORCID Grzegorz Oloś: 0000-0002-2096-980X

ORCID Agnieszka Dołhańczuk-Śródka: 0000-0002-9654-4111

## Abstract

At the basis of most risk assessments aimed at determining the long-term trends in changes in the activity of radionuclides in the environment and thus exposure to ionizing radiation, various concepts of half-lives are used, particularly: biological, environmental (ecological) and effective one. There is a clear lack of consensus on the terminological level and, more importantly, on the methodological level regarding the determination of these half-lives. This manifests by a divergent methodology and the existence of two main, but contradictory concepts. In the first, empirically determined half-life is referred to as environmental or biological. The effective half-life is extrapolated at a later stage after taking into account the physical decay. In the second concept, the effective half-life is the one empirically determined and the remaining half-lives can be extrapolated after correction for physical decay. As both concepts seems to be incompatible with each other, the aim of this work was to thoroughly analyze their theoretical assumptions, conditions, strengths and weaknesses and to define the conditions that must be met in order to enable comparison of the results obtained with the use of two separate concepts. In case of the environmental components the suggested approach is to empirically determine the effective half-life, and subsequently, to extrapolate the environmental and/or biological half-lives.

## 1. Introduction

The physical half-lives of radionuclides (RN), such as $^{137}Cs$, $^{90}Sr$, $^{241}Pu$, lasts dozens of years. As a result, the potential for isotopes to persist in the environment and risk of exposure to them by humans and/or biota is of great importance. The different half-lives, namely the effective ($T_{1/2eff}$), the environmental (ecological) ($T_{1/2env}$), and the biological one ($T_{1/2biol}$), are used for prediction of fate and behavior of RN in the environment. $T_{1/2eff}$ of a radioactive isotope can be used to describe changes of its activity in time in a given environmental component and is the result of its physical decay ($T_{1/2phys}$) and environmental processes, such as migration or elution which constitute environmental half-life ($T_{1/2env}$). Similarly, when calculating the $T_{1/2eff}$ in an organism, both physical decay and biological processes of excretion must be taken under consideration. For better clarity of this work the synonym for environmental half-life – “ecological half-life” will not be mentioned throughout the paper. However, it is understood that differences among $T_{1/2env}$ in biotic and abiotic environmental components exist as pointed out by Tagami and Uchida (2016)).

Different half-lives describing the loss of RN from a given component were described and clarified several decades ago, both in medical and environmental sciences (Landhal 1946, Naimark 1949, Comar 1955, Anderson 1957, Klement 1962, Broszkiewicz et al. 1963, French et al. 1965, Bergström 1967, Hanson & Eberhardt 1968, Kwan et al. 1977, 1984, and Boxenbaum & Battle 1995).

It is expected that such significant and well-established scientific terms would be described unambiguously and in a consistent manner. However, the available literature not only provides inconsistent nomenclature but also divergent methods of their calculation and interpretation. This, at least to some extent, has been pointed out decades ago (Lloyd & Rundo 1976).

Two divergent concepts are used widely in calculating $T_{1/2eff}$, $T_{1/2env}$ and $T_{1/2biol}$. According to the first concept, in a single compartment with first order kinetics, empirically determined changes in RN activity over time in a given component allows for the determination of $T_{1/2env}$ or $T_{1/2biol}$. The next step is to extrapolate $T_{1/2eff}$ after subsequent correction for physical decay. This approach can be found both in scientific publications (Pröhl & Hoffman 1992, Pröhl et al. 2006, Strebl & Tataruch 2007, Cevik and Celik 2009, Leclerc & Choi 2009, Lettner et al. 2009, Iurian et al. 2011, Brooks et al. 2015, Sunaga et al. 2015, Koivurova et al. 2015, Burger & Lichtscheidl 2018, Katengeza et al. 2020, Niizeki et al. 2020) and in the recommendations of organizations and scientific panels such as WHO, IAEA, and ENS (IAEA 1996, 2010, AMAP 2004, WHO 2008, ENS glossary).

The second concept assumes the opposite pattern of action. Due to the lack of precise data on how different processes effect the observed RN activity in the studied component, the empirically determined half-life is considered to be $T_{1/2eff}$. $T_{1/2env}$ or $T_{1/2biol}$ are subsequently extrapolated after a correction for physical decay, specific for a given RN (Ahman & Ahman 1994, Smith et al. 1999, Robison et al. 2003, Strandberg 2004, Hanslík et al. 2005, Zibold & Klemt 2005, Fesenko et al. 2009, Iwata et al. 2013, Paller et al. 2014, Rulík et al. 2014, Kapała et al. 2015, Merz et al. 2016, Steinhauser & Saey 2016, Corcho-Alvarado et al. 2016, Koarashi et al. 2016, Tagami and Uchida 2016, Savino et al. 2017, Matsuda et al. 2020, Kinase et al. 2020, Zotina et al. 2020, Hayes et al. 2020).

Since both concepts are based on the same equations, the precise description of the conditions that must be met in order to determine which half-life is obtained empirically in the first place needs to be given. Yet, such conditions are often not fulfilled or the same methodology of data collection is used in both concepts. Together with inconsistent terminology this seems to be a cause for which the discussed half-lives may be confusing, incomparable, or prone to bias, thus requiring an urgent need for attention and a constructive analysis. The aim of the study is to clarify and define the conditions allowing for the correct use of both concepts for the calculation of the mentioned half-lives.

## 2. Calculations

### 2.1 Dealing with terminological inconsistence

The analysis of the available literature revealed an inconsistency of terminology, particularly in $T_{1/2env}$ occurring in the form of the environmental loss half-life ($T_{ELH}$), environmental effective half-life ($T_{1/2env,eff}$) effective ecological half-life ($T_{eco,eff}$), ecological half-life ($T_{eco}$), and relative half-life ($T_r$) (Bagshaw & Brisbin 1985, Smith et al. 2000, Peles et al. 2002, Morita & Yoshida 2005, Chiaravalle et al. 2018, Nakanishi et al. 2019, Cevik & Celik 2009, Manaka et al. 2019, Yoshimura & Akama 2020). The effective half-life will be hereinafter referred to as $T_{1/2eff}$, the half-life which describes changes in the environmental component without physical decay as $T_{1/2env}$, and the half-life that describes changes of RN activity in a single organism without physical decay as $T_{1/2biol}$.

### 2.2 A coherent basis for both concepts

In a particular abiotic environmental component, the activity of a given RN present may differ due to environmental reasons (migration, dispersion, elution, etc.), and radioactive decay. In the case of a biotic component, such as the population of deer or moss within a particular area, RN activity will change depending on the factors that determine its activity in the environment, bioavailability, and the rate of excretion from the organisms of the studied species. In the case of a single organism, RN activity in its body and specific tissues will be determined by the RN physical decay, its intake, and the rate of its excretion. The result of these processes in all cases sets the rate and direction of changes in the activity of RN in time and is described as $T_{1/2eff}$.

If $T_{1/2phys}$ can be described by the use of an exponent in an exponential function according to the equation:

$$T_{1/2phys} = \frac{\ln(2)}{\lambda_{phys}} \qquad (1)$$

Then the environmental, biological and effective half-lives , with the use of $\lambda_{env}$, $\lambda_{biol,}$ and $\lambda_{eff}$ is respectively:

$$T_{1/2env} = \frac{\ln(2)}{\lambda_{env}} \quad or \quad T_{1/2biol} = \frac{\ln(2)}{\lambda_{biol}} \quad or \quad T_{1/2eff} = \frac{\ln(2)}{\lambda_{eff}} \qquad (2)$$

According to equation 3, the effective decay constant is the sum of physical and environmental decay. In other words, the observed changes in radionuclide activity must always be the sum of all processes influencing it's activity. If these processes follow the single first order kinetics model, then each of them can be described with a separate half-life, and so with different decay constant. The sum of the decay constants gives the effective decay constant. As a result, the inverse of effective half-life is the sum of the inverses of the physical and environmental half-lives (equation 4).

$$\lambda_{eff} = \lambda_{phys} + \lambda_{env} \qquad (3)$$

$$\frac{1}{T_{1/2eff}} = \frac{1}{T_{1/2phys}} + \frac{1}{T_{1/2env}} \qquad (4)$$

With the precise $T_{1/2phys}$ value and the value of one of the two other half-lives or its decay constants, one can calculate the missing half-life using simple mathematical transformations.

**2.3 Concept I:**

The crucial assumption of Concept I is that the observed changes in RN activity in the tested component, compiled on a semi-logarithmic activity vs. time plot, allows for empirical determination of $T_{1/2env}$ (in environmental component) or $T_{1/2biol}$ (in organism), according to equation 2 (IAEA 1996, 2010, WHO 2008). As stated by the IAEA and WHO, '*considering radioactive decay*' one can extrapolate $T_{1/2eff}$ by transforming equation 4 into:

$$T_{1/2eff} = \frac{T_{1/2phys} \cdot T_{1/2env}}{T_{1/2phys} + T_{1/2env}} \qquad (5)$$

The origins of using this concept dates back to the 1960s and 1970s, (Chipman 1967, Tuominen and Jaakkola 1973, Dapson and Kaplan 1975) and has been in use ever since (Pröhl & Hoffman 1992, AMAP 2004, Pröhl et al. 2006, Strebl & Tataruch 2007, Cevik and Celik 2009, Leclerc & Choi 2009, Iurian et al. 2011, Brooks et al 2015, Sunaga et al. 2015, Koivurova et al. 2015, Burger & Lichtscheidl 2018, Katengeza et al. 2020, Niizeki et al. 2020).

**2.4 Concept II:**

The second, widely used concept for $T_{1/2env}$, $T_{1/2biol}$ and $T_{1/2eff}$ calculations is based on an opposite assumption (Ahman & Ahman 1994, Robison et al. 2003, Strandberg 2004, Hanslík et al. 2005, Zibold & Klemt 2005, Fesenko et al. 2009, Iwata et al. 2013, Paller et al. 2014, Rulík et al. 2014, Kapała et al. 2015, Merz et al. 2016, Steinhauser & Saey 2016, Corcho-Alvarado et al. 2016, Koarashi et al. 2016, Tagami and Uchida 2016, Savino et al. 2017, Matsuda et al. 2020, Kinase et al. 2020, Zotina et al. 2020, Hayes et al. 2020). The crucial assumption of this concept is that one can empirically observe an effect of all, parallel processes that determine changes of RN activity in a studied component. There are two possible paths within Concept II for calculating $T_{1/2eff}$ and subsequently $T_{1/2env}$ or $T_{1/2biol.}$

The first path is based on $\lambda_{eff}$ derived from an exponential equation in the relationship between empirically determined RN activity vs. time in a semilogarithmic plot. Since the $T_{1/2phys}$ of all RNs is known, it is easy to calculate their $\lambda_{phys}$. Having both, the $\lambda_{eff}$ and $\lambda_{phys}$ one is able to calculate $\lambda_{env}$ or $\lambda_{biol}$ from equation 3, and $T_{1/2env}$ or $T_{1/2biol}$ using equation 2.

The second path within Concept II is based on a classical decay equation according to the one-compartment, single first order kinetics, which takes the form:

$$y = Ae^{-\lambda x} \qquad (6)$$

It is important to underline, that $\lambda_{eff}$ present in equation 6 and its transformations, reflects all possible processes by which RN changes its activity in the environment. Therefore, activity *N* after time *t* will amount to:

$$N_t = N_0 e^{-\lambda t} \qquad (7)$$

where:

*N* – radionuclide activity

$N_0$ – initial radionuclide activity

$\lambda$ – radionuclide effective decay constant in the studied component

*t* – time wherein *N* is determined

When deriving the equation for $T_{1/2eff}$, where RN activity equals half of its initial value $N_t = N_0/2$, after time t = $T_{1/2eff}$, it is necessary to transform equation 7 o the form of equation 2:

$$T_{1/2eff} = \frac{\ln(2)}{\lambda} \qquad (2)$$

Subsequently, if the value of $T_{1/2eff}$ is determined, and the specific value of $T_{1/2phys}$ is given, one can derive $T_{1/2env}$ by transforming equation 4 to the following form:

$$T_{1/2env} = \frac{T_{1/2phys} \cdot T_{1/2eff}}{T_{1/2phys} - T_{1/2eff}} \qquad (8)$$

When substituting $T_{1/2env}$ with $T_{1/2biol}$ and assuming that it is combined with monitoring changes of RN activity in the tissue or plasma of a particular organism, the mentioned equations are also fulfilled. Values of half-lives obtained using different equations: 8 or 2 within the Concept II are always the same.

## 3. Discussion

### 3.1 Discrepancies and ambiguities resulting from improper use of concepts I and II

Since all the equations in both concepts are correct, the core of the problem lies in the interpretation of what is empirically determined in the first place. Physical decay occurs regardless of any circumstances. Therefore, assuming that this process possibly, but not necessarily, contributes to the '*reduction of the content*' (IAEA 2010) of a particular RN in the studied component seems rather ambiguous, and will have serious consequences. In reality, decline in RN activity in a studied environmental component is an effect of parallel environmental processes and its' physical decay. If decline in RN activity is treated as an effect of environmental processes only, then subsequent correction for physical decay is redundant and will lead to significant misinterpretations.

To point out the possible consequences of incorrect use of Concept I, an example of contaminated sheep will be used, as given by IAEA (2010) in the chapter *Time dependence of radionuclide transfer factors*. It is stated that if a sheep contaminated with $^{137}Cs$ will be fed uncontaminated feed long enough, the level of $^{137}Cs$ in its body will decline "*at a rate determined by the biological half-life*". It seems that the decline of $^{137}Cs$ in sheep will not only be the consequence of biological processes (interpreted mainly as the exchange of caesium cation for $K^+$ (Burger & Lichtscheidl 2018)) but of all possible processes, including the physical decay ($T_{1/2phys}$) of the absorbed, but not yet excreted $^{137}Cs$. Indeed, in the absence of additional sources of $^{137}Cs$ in food with $T_{1/2biol} << T_{1/2phys}$ one can observe that $T_{1/2biol} \rightarrow T_{1/2eff}$, as a result of significant differences between the known values of these half-lives. This is the reason why in some cases the values of $T_{1/2eff}$ tend to the values of $T_{1/2env}$ or $T_{1/2biol}$ and why these terms are often treated interchangeably, as cited by Uchiyama (1978): "*…because physical half-life, 30 years, is long enough to allow the effective half-life to be the biological one*". However, the smaller the difference between $T_{1/2biol}$ / $T_{1/2env}$ and $T_{1/2phys}$ for a given RN, the greater the differences in the values of $T_{1/2eff}$ obtained from each concept. The hypothetical results obtained with the use of both discussed concepts are shown in Table 1, assuming that the change of RN activity was measured inside the studied component. The above analysis shows that the problem with the compatibility of the results obtained using concepts I and II has its source in how to define half-life calculated with equations 6, 7 and 2.

If the same equation (2) and the same data are used to calculate $T_{1/2env}$ or $T_{1/2eff}$ depending on the concept, the same value will be obtained but under a different name. This will have big consequences if the calculations are continued to determine the missing half-lives depending on the concept and chosen formula: 5 or 8. To highlight how big these differences may occur hypothetical data were presented in Table 1 for $^{137}Cs$ and $^{131}I$.

**Table 1. Differences in results of half-lives [years] obtained using the two discussed concepts**

| | | | Concept I | | Concept II | |
|---|---|---|---|---|---|---|
| **case** | **RN** | **$T_{1/2phys}$** [years] | **Hypothetical value of $T_{1/2biol}$ or $T_{1/2env}$ determined using equation no. 2** [years] | **$T_{1/2eff}$ determined using equation no. 5** [years] | **Hypothetical value of $T_{1/2eff}$ determined using equation no. 2** [years] | **$T_{1/2biol}$ or $T_{1/2env}$ determined using equation no. 8** [years] |
| 1 | $^{137}Cs$ | 30.17 | 0.1 | 0.1 | 0.1 | 0.1 |
| 2 | $^{137}Cs$ | 30.17 | 1.0 | 0.96 | 1.0 | 1.03 |
| 3 | $^{137}Cs$ | 30.17 | 30.0 | 15.09 | 30.0 | 5324.12 |
| 4 | $^{137}Cs$ | 30.17 | 30.17 | 20.18 | 30.17 | 0.0 |
| 5 | $^{137}Cs$ | 30.17 | 60.0 | 24.11 | 60 | - |
| 6 | $^{137}Cs$ | 30.17 | 1000 | 29.29 | 1000 | - |
| 7 | $^{131}I$ | 0.022 | 0.021 | 0.0107 | 0.021 | 0.462 |
| 8 | $^{131}I$ | 0.022 | 0.022 | 0.011 | 0.022 | 0.0 |

The value of $^{137}Cs$ $T_{1/2biol}$ for most of the living organisms (including sheep) is counted between a dozen of weeks to a few months (Twardock and Crackel, 1969, Mehli 1996, Kase 2012, CERN 2016). Thus, under laboratory conditions, use of both concepts will give close or indistinguishable results of $T_{1/2eff}$ (case 1 and 2). Yet, in a situation where the value of $T_{1/2env}$, calculated according to concept I, is equal to the value of $T_{1/2phys}$, the value of $T_{1/2eff}$ will be approximately two-thirds value of $T_{1/2phys}$ (case 4). Using the concept II, when the value of $T_{1/2eff}$ equals $T_{1/2phys}$, then $T_{1/2env}$ has no value - it does not exist, which seems rational. Moreover, according to concept I, when $T_{1/2env} > T_{1/2phys}$ (in case for any given radionuclide), then $T_{1/2eff}$ values will increase infinitely towards the $T_{1/2phys}$ value (case 5 and 6). This seems like a weakness of concept I. There are examples of environmental components characterized by high RN inflow. As a result, the activity of RN in such components may decrease slower than expected according to its physical decay or even increase over time when inflow outweighs the physical decay, e.g. wild boar populations in the Chernobyl exclusion zone (Steinhauser and Saey 2016). This limitation manifests itself when the value of the effective decay constant is greater than the physical decay constant for a given RN. In such a case, $T_{1/2env}$ or $T_{1/2biol}$ calculated using equation no. 8 will take a negative value, indicating that the “external” processes causes an inflow of the RN to the tested element (case 5 and 6).

Differences resulting from the use of Concept I, when it was necessary to use Concept II, based on the literature data, are presented in the form of a table in appendix 1.

**3.2 Reconciling the concepts I and II**

RN cannot "disappear" from the body through metabolism, so $T_{1/2biol}$ is based only on excretion processes. Thus, when using Concept I for $T_{1/2biol}$ calculation it needs to be estimated solely on RN activity just after it leaves the studied component, outside of that component. In the case of a single organism this means in all possible excretions (feces, urine, sweat, tears etc.). After this condition is fulfilled and the initial amount of RN is known, one can proceed with the use of Concept I as shown on the first scheme (Fig. 1). Only then, one can estimate the real value of $T_{1/2biol}$ not affected by physical decay. However, a limitation in the use of this approach is when $T_{1/2biol} > T_{1/2phys}$ (when biological processes take longer than physical to reduce the RN activity), because RN will decay to more than 50% of the initial dose before it is half excreted.

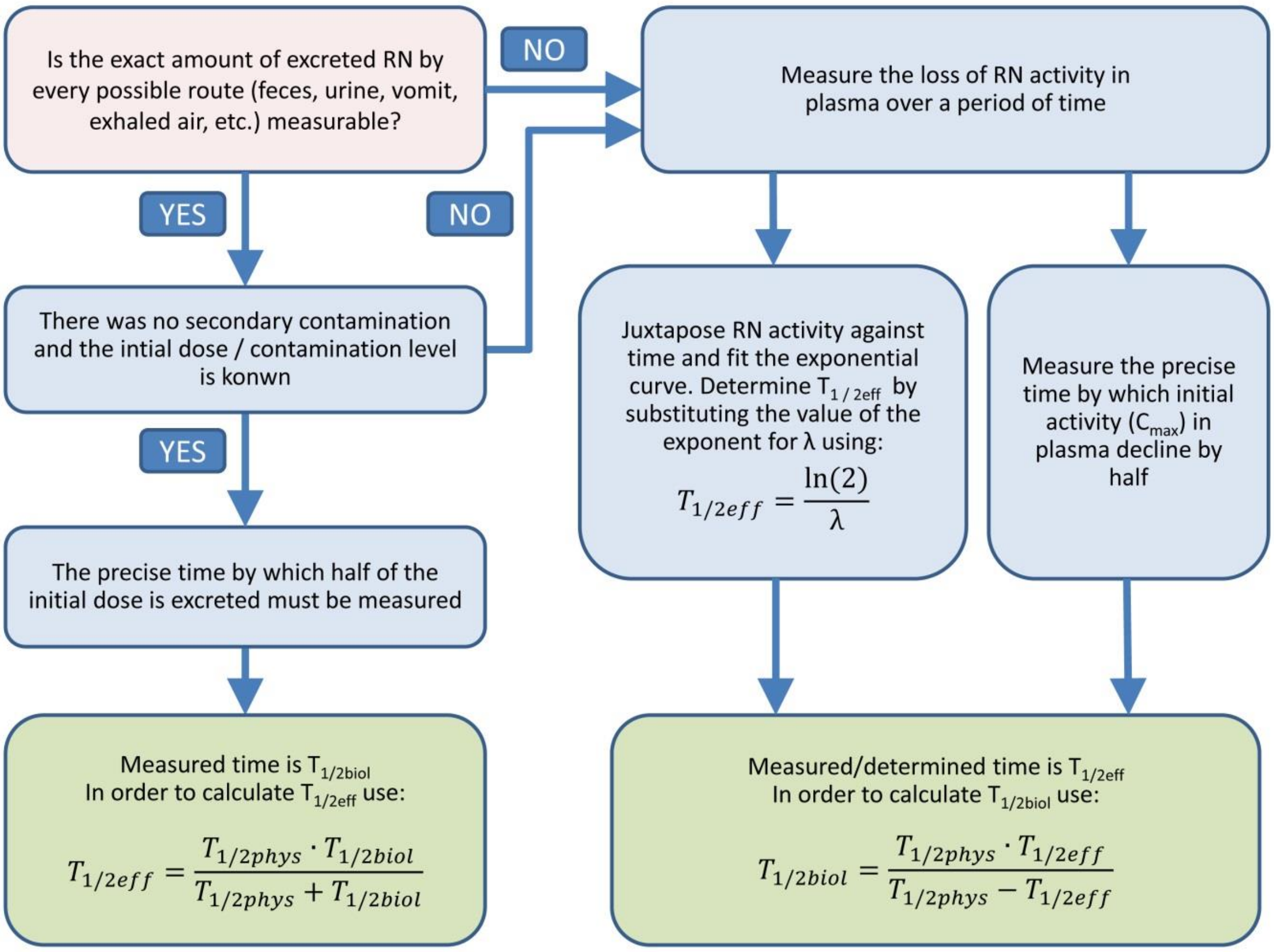

**Figure 1. Suggested procedure for $T_{1/2eff}$ and $T_{1/2biol}$ estimation in the case of an organism. The left column represents the use of Concept I and the right column the use of Concept II**

In the case where environmental components are studied instead of biological, strict procedures must be applied in order to achieve realistic $T_{1/2env}$ values with Concept I. However, in contrast to biological components, this would require a serious technical effort to measure the exact amounts of RN that has left the studied component on all possible routes. Only then, one can proceed as shown on the second scheme (Fig. 2). Moreover, the same limitations as for the $T_{1/2biol}$ calculated with a use of Concept I will also manifest (when $T_{1/2env} > T_{1/2phys}$ for a specific RN).

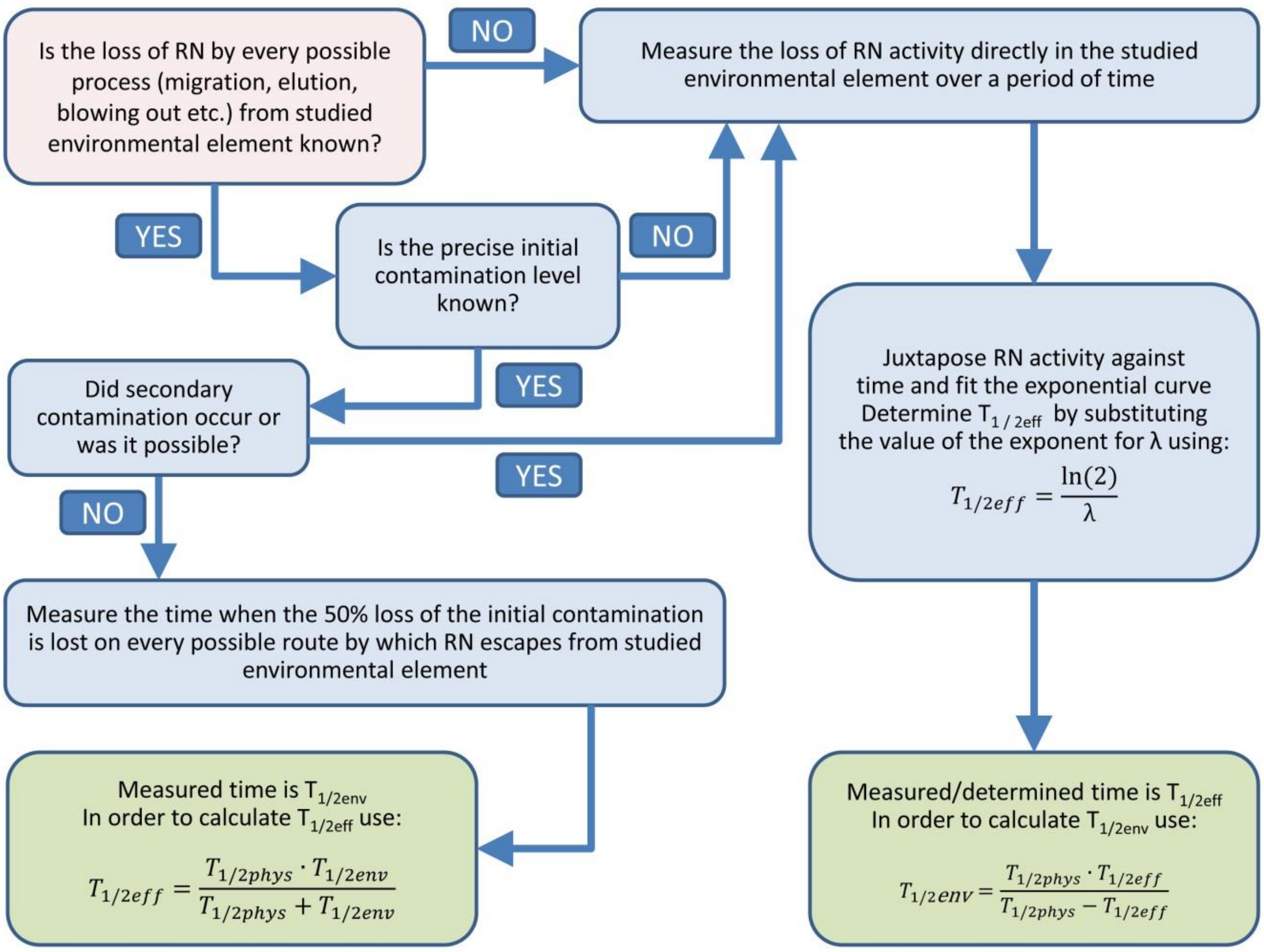

**Figure 2. Suggested procedure for $T_{1/2eff}$ and $T_{1/2env}$ estimation in case of an environmental element (including biota)**

The technical information regarding measurement of RN activity outside of the tested environmental or biological component seems to be obligatory each time Concept I is used. However, this information has not been given in all cited publications prompting the use of Concept I. This creates a significant problem, as this concept is suggested by well-established sources like Pharmacopeia of WHO (WHO 2008) or the European Nuclear Society (ENS glossary). Given exemplary values of $T_{1/2biol}$ and $T_{1/2eff}$ in ENS glossary clearly indicate the use of Concept I. Moreover, this way of calculating the discussed half-lives has steeped deeply in popular scientific resources. Without providing a detailed methodology on how the given values were estimated in a first place, results acquired with the use of Concept I are doubtful.

Past work exists in which the complete methodology of half-live calculation is not given or precisely described (Machart et al. 2007, Outola et al. 2009, Petraglia et al. 2020, Gaines et al. 2021, Yu et al. 2021). In such cases, it is not possible to determine which concept has been used. As a result, comparisons between these studies with work that does include a complete methodology will be challenging.

## 4. Conclusions

The heart of the problem seems to be the interpretation of which half-life is determined empirically in an organism or environmental component, and which is extrapolated. Changes in the activity of a given RN in the studied component are always a result of all possible parallel processes, including its radioactive decay. All together they determine the $T_{1/2eff}$ of a given RN. If so, interpreting the empirically determined half-lives inside the tested medium as $T_{1/2env}$ or $T_{1/2biol}$, with subsequent correction for physical decay leads to serious misinterpretations. The discrepancies resulting from an inappropriate use of Concept I, along with inconsistent terminology, may be responsible for the difficulties in comparing and interpreting the discussed half-lives. In addition, this may create doubt in any risk assessment based on half-lives obtained this way. The inappropriate use of Concept I based on the measurements conducted inside the tested component will always be subject to a certain bias. The only way for a precise estimation of $T_{1/2env}$ or $T_{1/2biol}$ in a first place using the Concept I, is by a detailed measurement of RN activity that has left the tested component, but this must be supported with a detailed methodology. When this condition is met, the precision of both concepts will depend on the object of the study and technical abilities of measuring the loss of radionuclides from the studied component. In contrast, determination of $T_{1/2eff}$ using Concept II, on the basis of changes in RN activity over time in the tested component, with the subsequent extrapolation of $T_{1/2biol}$ or $T_{1/2env}$ seems to be technically simpler, more precise, and less expensive, especially in environmental studies. Its use, along with a more precise nomenclature, should facilitate future research of this type and comparisons between results.

## Declaration of interest statement

The authors declare that there is no conflict of interest.

This work received no specific grant from any funding agency, commercial or not-for-profit sectors.

## Acknowledgments

We would like to express our gratitude to Natasha Ng for language help provided in writing of this manuscript.